\documentclass[a4paper,11pt]{article}
\usepackage{jcappub} 

\usepackage{amsmath, amssymb, amsfonts, amsbsy, amstext, amsthm}
\usepackage{graphicx}
\usepackage{tabularx}
\usepackage[toc,page]{appendix}
\usepackage{siunitx}
\usepackage{mathtools}
\usepackage{multirow}
\usepackage[usenames,dvipsnames,table,xcdraw]{xcolor}
\usepackage{aas_macros}
\usepackage{xspace}
\usepackage{ulem}
\usepackage{mathrsfs}

\usepackage{tabularray}
\UseTblrLibrary{booktabs}

\title{\boldmath Improving Constraints on Models Addressing the Hubble Tension with CMB Delensing}

\newcommand{\smu}{Department of Physics,
Southern Methodist University,
Dallas, TX 75275, USA}

\author{Joshua Ange}
\author{and Joel Meyers}
\affiliation{\smu}

\emailAdd{jwange@mail.smu.edu}

\abstract{The Hubble Tension is a well-known issue  in modern cosmology that refers to the apparent disagreement in inferences of the Hubble constant $H_0$ as found through low-redshift observations and those derived from the $\Lambda$CDM model utilizing early universe observations. Several extensions to $\Lambda$CDM have been proposed to address the Hubble Tension that involve new ingredients or dynamics in the early universe. Reversing the effects of gravitational lensing on cosmic microwave background (CMB) maps produces sharper acoustic peaks in power spectra and allows for tighter constraints on cosmological parameters. We investigate the efficacy of CMB delensing for improving the constraints on parameters used in extensions of the $\Lambda$CDM model that are aimed at resolving the Hubble Tension (such as varying fundamental constants, contributions from early dark energy, and self-interacting dark radiation). We use Fisher forecasting to predict the expected constraints with and without this delensing procedure. We demonstrate that CMB delensing improves constraints on $H_0$ by $\sim$ 20\% for viable models and significantly improves constraints on parameters across the board in the low-noise regime.}

\begin{document}
\maketitle
\flushbottom

\section{Introduction}
\label{sec:intro}

The cosmic microwave background (CMB) has been instrumental in shaping our modern understanding of the evolution of the universe. With the increase of precision CMB observations available, temperature and polarization anisotropies have become more finely measured, leading to more precise inferences of the contents and history of the universe. The physical mechanisms driving the formation of the CMB and its anisotropies are well understood, leading to a robust prediction for the form of the angular power spectra for any specific cosmological model. Based on ever-improving measurements of the CMB anisotropies, the cosmological parameters driving these models have become more tightly constrained~\cite{Planck:2018vyg}.

Acoustic oscillations in the primordial plasma prior to recombination imprinted a series of peaks and troughs in CMB power spectra. The expansion history and plasma properties before recombination set the physical scale of the sound horizon, the expansion history after recombination sets the distance to the surface of last scattering, and the combination of the two sets determine the angular scale of the sound horizon at recombination. Measurements of the CMB acoustic peak positions allow for a direct measurement of the angular size of the sound horizon at recombination.  Additionally, the relative peak heights allow for an inference of the parameters that determine the scale of the sound horizon within a given cosmological model. Combining this information allows for an inference of the integrated expansion since recombination, including in particular the present expansion rate, the Hubble constant $H_0$ \cite{Knox:2019rjx}. While this is a useful qualitative picture for how constraints are derived from CMB observation, in practice, the parameter constraints for any particular cosmological model are obtained from a full likelihood analysis of the model given the CMB data.

The past decade has seen an ever-growing tension in the value of $H_0$ as determined from local measurements \cite{Riess:2020fzl} and as inferred from early universe observations like the CMB \cite{Planck:2018vyg} within $\Lambda$CDM cosmology. Today, this ``Hubble Tension" is driven primarily from the SH0ES (Supernovae and $H_0$ for the Equation of State of dark energy)  collaboration's cosmic distance ladder measurements and the Planck collaboration's CMB observations \cite{Schoneberg:2021qvd}. This apparent discrepancy may be due to some unaccounted systematic error in one or more of the measurements, or it may point toward some physics that is left out of our current cosmological models. Many attempts at resolving the tension take the route of new physics~\cite{DiValentino:2021izs}. In this work, we have found that implementing CMB delensing leads to significantly improved constraints on the parameters defining models aimed at resolving the Hubble tension, and these improved constraints remain valid even if the Hubble Tension does not end up directly involving new physics.

Between the CMB surface of last scattering and our telescopes, large scale structure gravitationally deflects incoming photons, producing non-stationary statistics of the CMB fluctuations. At the level of the power spectra, this lensing effect smooths acoustic peaks, making their angular scale more difficult to measure and thus makes some parameter constraints less precise than they otherwise would be~\cite{Lewis:2006fu}.  Current measurements allow for a detection of the effects of CMB lensing at a significance of 40$\sigma$ with Planck data~\cite{Planck:2018lbu}, with similar precision from  ACT~\cite{ACT:2023dou} and SPT~\cite{Wu:2019hek}.For future low-noise CMB data, such as that expected from Simons Observatory \cite{SimonsObservatory:2018koc} and CMB-S4 \cite{CMB-S4:2016ple,Abazajian:2019eic}, lensing will be measured at much higher significance, and peak smoothing will hinder some parameter constraints.  Fortunately, the off-diagonal mode correlations induced by lensing allow for the reconstruction of a lensing potential map~\cite{Hu:2001kj,Okamoto:2003zw}, which can be used to reverse the effects~\cite{Hirata:2002jy,Hirata:2003ka,Smith:2010gu}. Delensing temperature and polarization maps generates spectra with sharper acoustic peaks, more prominent damping tails, and other clearer features, thereby allowing for tighter parameter constraints~\cite{Green:2016cjr,Hotinli:2021umk}.

Here, we provide a quantitative look at the improvements made possible by CMB delensing, specifically in regard models aimed at resolving the Hubble Tension. Many proposed solutions to the tension take the form of extensions to the $\Lambda$CDM model. By introducing new physics that modify the sound horizon, a shift can be made in the inferred value of $H_0$, aligning it more closely with local measurements \cite{Schoneberg:2021qvd}. We have considered three broad categories of early universe solutions to the Hubble Tension: those involving a variation of fundamental constants at early times, those involving early dark energy, and those involving self-interacting dark radiation. We generate $TT$, $TE$, $EE$, and $\phi\phi$ power spectra from the best-fit values of model parameters that most ease the Hubble Tension, we implement iterative delensing to produce delensed spectra for these models, and we perform Fisher forecasts to determine the expected parameter constraints from upcoming CMB observations. We show that CMB delensing leads to significantly tighter forecasted constraints on models aimed at resolving the Hubble Tension.

\begin{figure}[t!]
    \centering
    \includegraphics[width=\columnwidth]{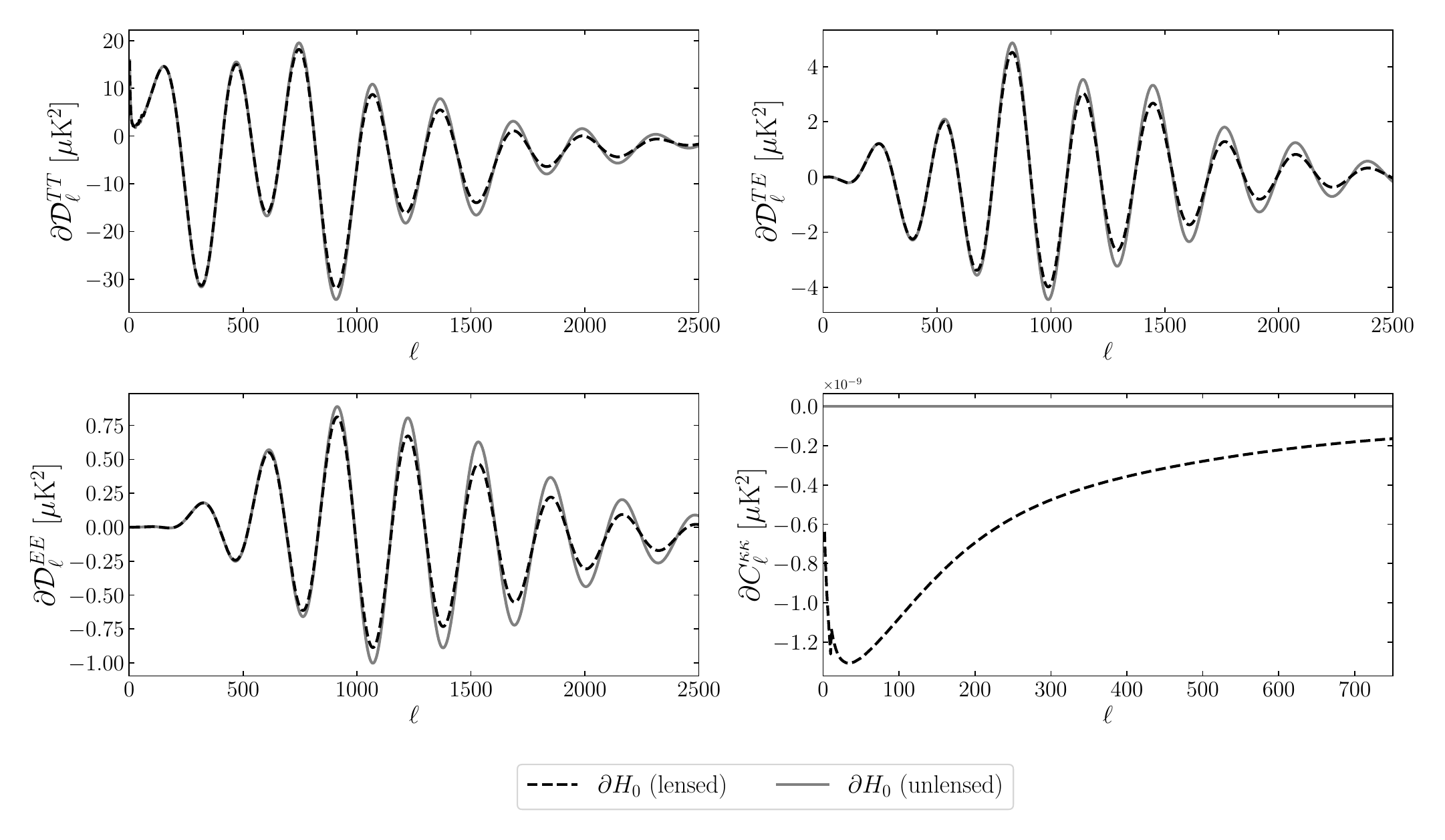}
    \caption{Derivatives of CMB power spectra ($TT$, $TE$, and $EE$) and lensing power spectrum ($\kappa\kappa$) from $\Lambda$CDM with respect to $H_0$. In this figure and below, $H_0$ is taken to be measured in units of \unit{km/s/Mpc}. Acoustic features like peak positions allow us to infer the angular size of the sound horizon \cite{1997mba..conf..333H}, so by shifting $D_A^*$, a phase shift is made on power spectra features. Note the sharper peaks of the unlensed spectra lead to larger derivatives with respect to $H_0$.}
    \label{fig:spectraDeriv_H0}
\end{figure}

In Section~\ref{sec:Hubble_tension}, we outline the Hubble Tension and how $H_0$ is inferred from the CMB. We briefly overview how iterative delensing works and affects constraining power derived from CMB observations in Section~\ref{sec:delensing}, and we also provide the details of our forecasts there. In Section~\ref{sec:results}, we show the degree to which CMB delensing improves constraints across the proposed solutions. Section~\ref{sec:conclusion} concludes.

\section{The Hubble Tension}
\label{sec:Hubble_tension}

There is a growing tension in the value of the current expansion rate of the universe, the Hubble Constant $H_0$, as inferred from early universe observations and from local measurements. The tightest local measurements utilize the Cepheid-calibrated cosmic distance ladder, and the SH0ES team provides a constraint of 
$H_0 = 73.2 \pm 1.3~\unit{km/s/Mpc}$ \cite{Riess:2020fzl}. Alternatively, early universe observations like those of the CMB can be used to infer the Hubble constant within a particular cosmological model, and the Planck collaboration gives
$H_0 = 67.4 \pm 0.5~\unit{km/s/Mpc}$ in $\Lambda$CDM cosmology \cite{Planck:2018vyg}. 

To measure the Hubble constant from early universe observations, one evaluates the likelihood of parameter values given some set of measurements to match theoretical and observational findings. It is enlightening to consider qualitatively how the CMB power spectra can be used to constrain $H_0$~\cite{Knox:2019rjx}.  This involves calculating the sound horizon $r_s^*$ at the CMB and its angular size $\theta_s^*$ to determine the comoving angular diameter distance to the surface of last scattering $D_A^* = \frac{r_s^*}{\theta_s^*}$. The size of the comoving sound horizon at the CMB last scattering surface is
\begin{equation} \label{eqn:sound_horizon}
    r_s^* = \int_{0}^{t_*} \frac{dt}{a(t)} c_s(t) = \int_{z_*}^{\infty} \frac{dz}{H(z)} c_s(t)\text{,}
\end{equation}
where $t_*$ and $z_*$ refer to the time and redshift values associated with the CMB surface of last scattering. The sound horizon is dependent on the time of recombination, the sound speed in the primordial plasma $c_s(t)$, and the Hubble parameter $H(z)$ prior to recombination (which is dependent on the energy budget of the universe). The Fourier modes of density perturbations in the primordial plasma underwent damped and driven oscillations, and the angular size of the sound horizon $\theta_s^*$ can be directly observed from the peak spacing of the Fourier modes in CMB spectra \cite{1997mba..conf..333H}. The comoving angular diameter distance to the surface of last scattering can then be determined from $r_s^*$ and $\theta_s^*$. The comoving angular diameter distance to the surface of last scattering is given by $D_A^* = \int_0^{z_*} \frac{dz}{H(z)}$, thereby allowing a determination of $H_0$ for a given cosmological model.

The tension in the $H_0$ inference is not restricted to the comparison of local measurements with CMB data.  Constraints on $H_0$ derived from the combination of baryon acoustic oscillation data and measurements of the primordial light element abundances agree with the inferences from the CMB and show a tension with the SH0ES constraint~\cite{Addison:2017fdm,Cuceu:2019for,Schoneberg:2019wmt}, with a recent analysis of primordial deuterium abundance and SDSS DR16 data giving $H_0 = 68.3 \pm 0.7~\unit{km/s/Mpc}$~\cite{Schoneberg:2022ggi}.  The basic idea of these constraints is that baryon acoustic oscillation measurements provide joint constraints on $H_0$, $\Omega_m$, and $\omega_b$, and adding the tight constraint on $\omega_b$ obtained from observing the abundance of deuterium produced during big bang nucleosynthesis allows for a precise inference of $H_0$.  This combination of data is independent of CMB observations, indicating that the Hubble tension is very unlikely to be due to a systematic error in the CMB observations.

Similarly, local measurements other than those based on Cepheid-calibrated supernovae can be used to constrain $H_0$~\cite{Freedman:2021ahq}.  An approach based on the distance ladder calibrated with the tip of the red giant branch rather than with Cepheids provides a measurement of $H_0$ that agrees with both the early universe and SH0ES values~\cite{Freedman:2019jwv}.  Inferences of $H_0$ from strong lensing time delays tend to agree more closely with the SH0ES measurement, though the best fit value and size of the confidence interval depends on modeling and analysis choices~\cite{Wong:2019kwg,Birrer:2020tax,Birrer:2020jyr,Shajib:2023uig}.


The persistent discrepancy between local and early universe measurements of $H_0$ has prompted a wide range proposals for new cosmological models aimed at resolving the tension~\cite{DiValentino:2021izs,Schoneberg:2021qvd}. Proposed solutions to the tension often take the form of extensions to the $\Lambda$CDM model that alter one or more aspects of the cosmological history and shift our inference of $H_0$. This typically involves modifying the sound horizon of the CMB at last scattering with additional energy contributions or by modifying one of the other assumptions built into the $\Lambda$CDM model (such as the time of recombination). Such changes modify early universe dynamics and impact the inference of $H_0$ (see Fig.~\ref{fig:spectraDeriv_H0}). These types of models can often shift the inferred value of $H_0$ to be closer to the value obtained by SH0ES, but the changes to the cosmological model also impart other changes to the CMB anisotropies, which leads to constraints on these proposed solutions.

One possible resolution to the Hubble tension is that something is wrong with one or more of the measurements involved in obtaining the differing values of $H_0$ or that the uncertainties in those inferences have been underestimated.  Our goal for this paper is not to assess whether the Hubble tension requires new physics or to evaluate any particular model.  Rather, we demonstrate that for a broad class of models aimed at resolving the Hubble tension, CMB delensing provides a means by which we can obtain tighter constraints on the parameters defining those models.

\section{Delensing and Forecasting}
\label{sec:delensing}

Gravitational deflection of CMB photons by the cosmological structure intervening between the surface of last scattering and our telescopes distorts our view of CMB anisotropies~\cite{Lewis:2006fu}.  This deflection can be both a help and a hindrance, as it allows for a means to determine the distribution of matter, but also distorts the primary anisotropies of the CMB.  Lensing remaps the CMB temperature and polarization along some direction $\boldsymbol{n}$ by a deflection angle $\boldsymbol{d}(\boldsymbol{n})$ such that $T^\mathrm{len}(\boldsymbol{n}) = T^\mathrm{unl}(\boldsymbol{n} + \boldsymbol{d}(\boldsymbol{n}))$, and similarly for the Stokes $Q$ and $U$ parameters defining linear polarization.  Gravitational lensing modifies the statistics of CMB anisotropies, inducing correlations between fluctuations of different angluar sizes.  By considering quadratic combinations of CMB fields (such as the minimum variance quadratic estimator), a map of gravitational deflection can be estimated, thereby providing a map of the integrated matter density throughout the universe~\cite{Hu:2001kj,Okamoto:2003zw}.

Delensing aims to utilize this understanding of the effects of gravitational lensing from measured CMB maps (such as $T^\text{obs}$) to estimate lensing deflection $\boldsymbol{d}^\text{obs}$ and to reverse the effects of lensing. 
Iterative delensing is the process whereby one alternates lensing reconstruction and delensing, obtaining a better lensing estimate and more thoroughly delensed CMB maps at each step~\cite{Hirata:2002jy,Hirata:2003ka,Smith:2010gu,Hotinli:2021umk}.  For this paper, we utilize the publicly available code \texttt{CLASS\_delens}\footnote{\url{https://github.com/selimhotinli/class_delens}} as described in Ref.~\cite{Hotinli:2021umk} (a modified version of the Boltzmann solver \texttt{CLASS}~\cite{Blas:2011rf}), which computes the power spectra expected from performing an iterative delensing procedure to CMB maps.


Our forecasts were carried out using the \texttt{FisherLens} code\footnote{\url{https://github.com/ctrendafilova/FisherLens}}, described in Ref.~\cite{Hotinli:2021umk}. We conduct forecasts for a range of noise levels, and for all configurations, we utilized a beam size of $1.4\,\unit{arcmin}$, sky coverage of $f_\text{sky} = 0.5$, and a range of 30 to 5000 in $\ell$ in our analysis (though we restricted the range for $TT$ spectra to 30 to 3000 in $\ell$ since astrophysical foregrounds are expected to contaminate smaller scales). We take noise in polarization to be given by $\Delta_P = \sqrt{2}\Delta_T$, as is expected from fully polarized detectors. For each model, we chose fiducial values for the cosmological parameters that best reduced the Hubble Tension according to analyses of those models in the references given below. We impose a prior constraint on the optical depth to reionization of $\sigma(\tau) = 0.007$.

Our forecasted constraints include the effects of lensing-induced non-Gaussian power spectrum covariances as described in Refs.~\cite{Green:2016cjr,Hotinli:2021umk}.
In conducting forecasts for some models (particularly those involving varying fundamental constants), constraints for some parameters and configurations derived from delensed spectra with lensing-induced non-Gaussian covariances included were marginally smaller than those derived from delensed spectra with purely Gaussian covariances. We expect non-Gaussian covariance to weaken the constraining power~\cite{Hu:2001fb,Benoit-Levy:2012dqi,Schmittfull:2013uea,Peloton:2016kbw,Green:2016cjr,Hotinli:2021umk}, so, in these cases, we reported the larger errors neglecting the non-Gaussian covariance. This discrepancy is likely due to our approximation of the lensing-induced non-Gaussian covariance (see the technical details in Refs.~\cite{Green:2016cjr,Hotinli:2021umk}). 
We expect that, since delensing acts to reduce the non-Gaussian covariance for well-measured modes, the error introduced by neglecting it is small. We leave a more complete and robust treatment of the lensing-induced non-Gaussian covariance for future work. 

We carry out forecasts both with and without the Baryon Acoustic Oscillation (BAO) information anticipated from DESI (Dark Energy Spectroscopic Instrument)~\cite{Font-Ribera:2013rwa}. We constructed a Fisher matrix for BAO observables following the prescription given in Ref.~\cite{Allison:2015qca}. Forecasts with BAO included simply add this Fisher matrix to the one obtained from the CMB.  As shown below, the improvement in constraints from CMB delensing is smaller, though still non-trivial, when BAO information is included.

\section{Results}
\label{sec:results}

As the precision of both local distance ladder measurements and $\Lambda$CDM inferences have improved, and the Hubble Tension has grown more statistically significant, a number of solutions have been proposed. We consider models that were found to perform well for reducing the tension according to the survey of models described in Ref.~\cite{Schoneberg:2021qvd}. In that work, a wide variety of models and modifications were investigated, involving early- and late- universe physics, and we focus on representative examples of the three broad classes of early universe solutions: varying fundamental constants, early dark energy, and self-interacting dark radiation.

\subsection{Varying Fundamental Constants}
\label{subsec:varyingconstants}

\begin{figure}[t!]
    \centering
    \includegraphics[width=\columnwidth]{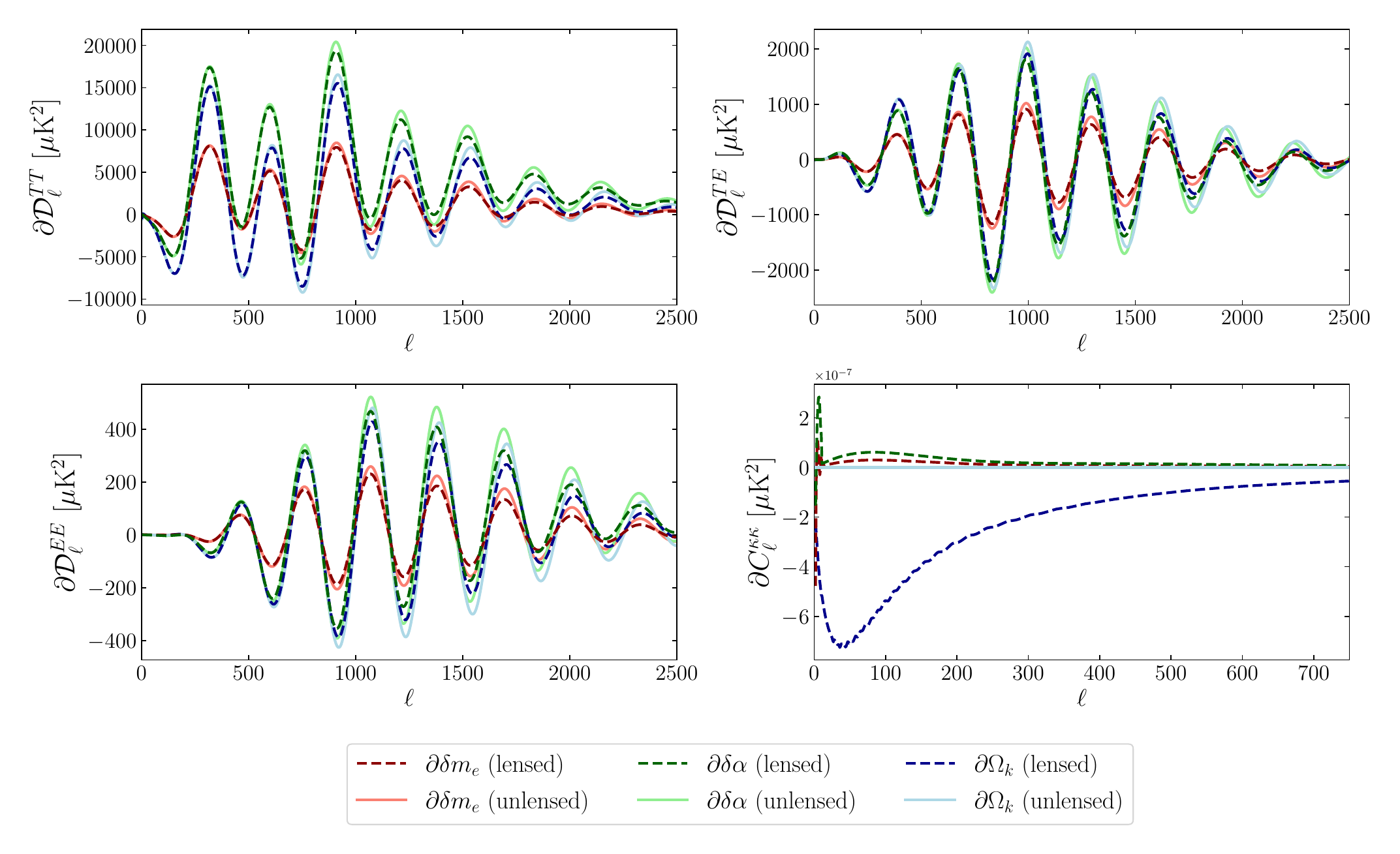}
    \caption{Derivatives of CMB power spectra ($TT$, $TE$, and $EE$) and lensing power spectrum ($\kappa\kappa$) with respect to the variation of the electron mass $\delta m_e$, the variation of the fine structure constant $\delta \alpha$, and the spatial curvature parameter $\Omega_k$. The larger derivatives for unlensed (and therefore delensed) spectra compared to lensed spectra indicate CMB peaks and features could be better isolated and more Fisher information could be obtained, so tighter constraints could be found with delensing.}
    \label{fig:spectraDeriv_varyingConstants}
\end{figure}
\begin{figure}[t!]
    \centering
    \includegraphics[height=4cm]{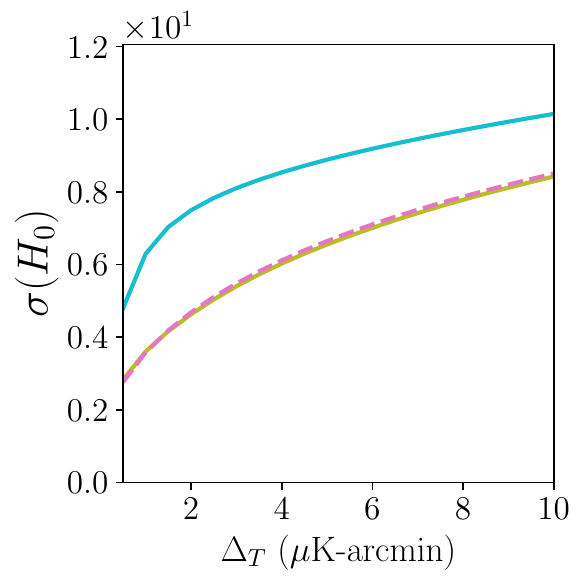}
    \includegraphics[height=4cm]{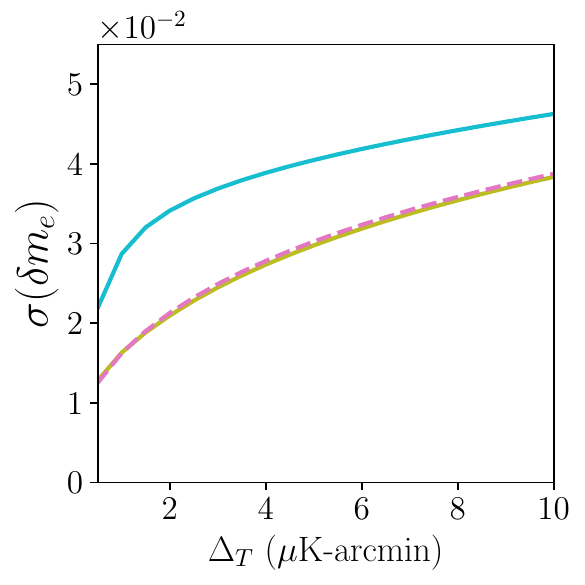}
    \includegraphics[height=4cm]{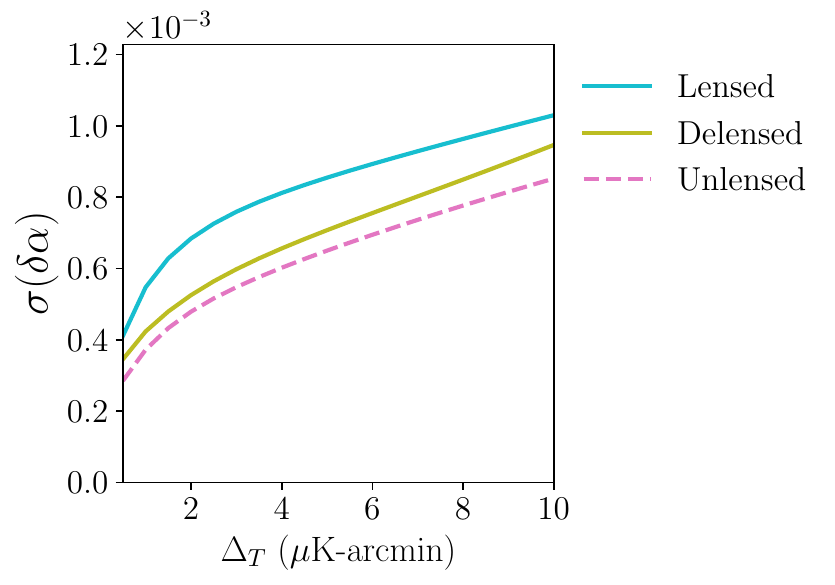}
    \caption{Forecasted constraints on $H_0$, $\delta m_e$, and $\delta \alpha$ in an 8-parameter $\Lambda\text{CDM} + \delta m_e + \delta \alpha$ cosmology. For $\delta m_e$ and degenerate parameters like $H_0$, using delensed spectra provides significantly tighter constraints compared to lensed spectra. 
    }
    \label{fig:constraints_varyingConstants}
\end{figure}
\begin{figure}
    \centering
    \includegraphics[width=0.4\columnwidth]{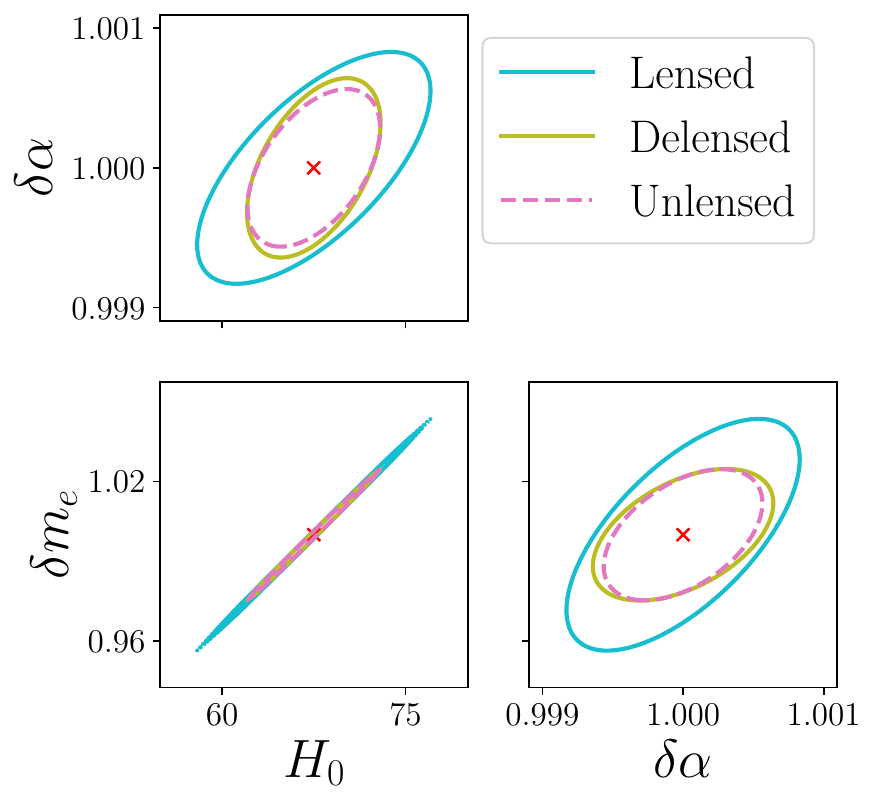}
    \caption{Forecasted errors and degeneracy of parameters of $\Lambda\text{CDM} + \delta m_e + \delta \alpha$ cosmology at $1~\unit{\mu K}$-arcmin. 
    }
    \label{fig:contours_varyingConstants}
\end{figure}

One class of proposed models that is effective at reducing the Hubble Tension comes in the form of varying fundamental constants, particularly those affecting recombination. By positing a time-varying electron mass and fine structure constant, the energy levels of Hydrogen atoms, the Thomson scattering cross-section, and other physics of the CMB at recombination are shifted \cite{Sekiguchi:2020teg}. These changes affect the energy and temperature at which recombination occurs, indicating a strong degeneracy of those parameters and the time of recombination $t_*$.
This leads to a change in the comoving size of the sound horizon, meaning the inference of $H_0$ can be made more compatible with local measurements \cite{Franchino-Vinas:2021nsf, Hart:2019dxi}. In these models, it is assumed that the fundamental constants shift to their currently measured values sometime after recombination, but well before reionization.  Observable impacts of this model should be visible in CMB power spectra, as a shifted time of recombination implies oscillations in the primordial plasma `froze' at a different point in their development, thereby shifting acoustic peaks (see Fig.~\ref{fig:spectraDeriv_varyingConstants}). CMB delensing produces sharper peak locations, meaning tighter constraints for these parameters could be found.

We define the time-varying parameters
\begin{equation}
	\delta m_e = \frac{m_{e,\text{early}}}{m_{e,\text{obs}}}\text{, \hspace{0.1cm}}\delta \alpha = \frac{\alpha_{\text{early}}}{\alpha_{\text{obs}}}
\end{equation}
where the early values of these parameters are time-independent throughout recombination, but change to their current, observed values at $z = 50$ following the treatment in Ref.~\cite{Hart:2017ndk}.

In Fig.~\ref{fig:constraints_varyingConstants} we show the results for 8-parameter forecasts where we allow for varying, non-unity values of $\delta m_e$ and $\delta \alpha$ in addition to the $\Lambda$CDM parameters. From left to right, the panels show constraints for $H_0$, $\delta m_e$, and $\delta \alpha$, respectively, across a range of CMB noise levels. At decreasing noise levels, delensing provides a significant improvement in the constraints. The improvement from lensed to delensed constraints is shown in Table~\ref{tbl:improvement_varyingConstants} for various combinations of fixed parameters.

\begin{table}[t!]
\centering
\begin{booktabs}{
  colspec = {lcccccc},
  cell{1}{2,5} = {c=3}{c}, 
  cell{1}{1} = {c=1}{c},
}
\toprule
  Model                                            &  With BAO & &                                    & Without BAO & & \\
\midrule
                                                   &  $H_0$         & $\delta\alpha$ & $\delta m_e  $ &  $H_0$         & $\delta\alpha$ & $\delta m_e  $ \\
\cmidrule[lr]{2-4}\cmidrule[lr]{5-7}
$\Lambda\text{CDM}+\delta\alpha+\delta m_e$        & 1.9            &  10.8          & 5.8            & 30.6           & 16.3           & 30.9           \\
$\Lambda\text{CDM} + \delta \alpha$                & 5.8            &  9.0           &                & 9.1            & 10.9           &                \\
$\Lambda\text{CDM} + \delta m_e$                   & 2.1            &                & 3.9            & 26.1           &                & 26.4           \\
$\Lambda\text{CDM}$                                & 2.2            &                &                & 4.6            &                &                \\
\bottomrule
\end{booktabs}
\caption{Table displaying percent improvement in error from delensing in the case of varying fundamental constants. This is computed with $100\times(1- \frac{\sigma_\text{del}}{\sigma_\text{len}})$ at $1 ~\unit{\mu K}$-arcmin (the approximate noise level of CMB-S4). Empty cells indicate parameters fixed at their fiducial values.
}
\label{tbl:improvement_varyingConstants}
\end{table}

In Fig.~\ref{fig:contours_varyingConstants}, we show the forecasted uncertainty and degeneracy of parameters of the $\Lambda\text{CDM} + \delta m_e + \delta \alpha$ model (without BAO) before and after delensing. Delensing provides a significant tightening of constraint contours for all parameters of interest. Additionally, there is a strong degeneracy between $\delta m_e$ and $H_0$ which is partially broken by delensing, as can be seen in the second and fourth rows of Table~\ref{tbl:improvement_varyingConstants}.

Note, in the middle panel of Fig.~\ref{fig:constraints_varyingConstants} (and indeed for all constraints involving $\delta m_e$ or degenerate parameters), the delensed spectra give tighter constraints than the unlensed spectra. In the limit that there is perfect knowledge of the lensing spectrum, this would be senseless, but with  noisy lensing reconstruction (as computed in our analysis), this is a valid result. If one considers a parameter for which the observable impact is entirely contained in the lensing (and not the primary CMB), without lensing reconstruction tighter constraints would undoubtedly be found from lensed spectra rather than unlensed spectra. With a noisy, sub-optimal lensing reconstruction, one would still find tighter constraints from the lensed spectra (and noisy reconstruction) than the unlensed spectra (and noisy reconstruction), as the poorly reconstructed components of the lensing spectrum still influence the lensed CMB data, while no information for the parameter is found in the unlensed data. Likewise, the delensed spectra should produce tighter constraints than the unlensed spectra, as only the lensing modes that are well-measured are removed. In the opposite limit, with a parameter that affects the primary CMB and not the lensing spectrum, the unlensed CMB spectra would outperform the lensed and delensed spectra regardless of reconstruction fidelity. For parameters that impact both the primary CMB and lensing spectrum, it does not necessarily hold that the unlensed spectra error should always be smaller the delensed spectra error, so delensing actually produces tighter constraints than the unlensed CMB would. More generally, when considering constraints on several cosmological parameters, with varied and partially degenerate impacts on the primary and lensing spectra, delensing should always improve constraints compared to analyses using lensed spectra, and in some cases delensed spectra may provide tighter constraints than unlensed spectra, so long as the reconstructed lensing spectrum is included in the analysis.

\subsubsection{Spatial Curvature}
\label{subsubsec:SpatialCurvature}

\begin{figure}[t!]
    \centering
    \includegraphics[height=3.75cm]{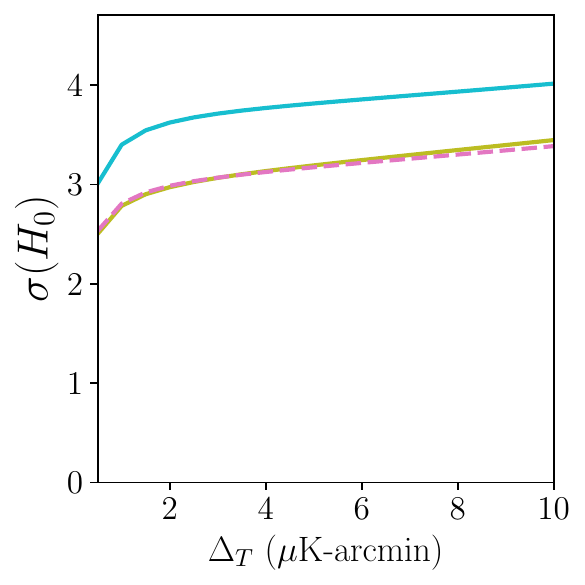}
    \includegraphics[height=4cm]{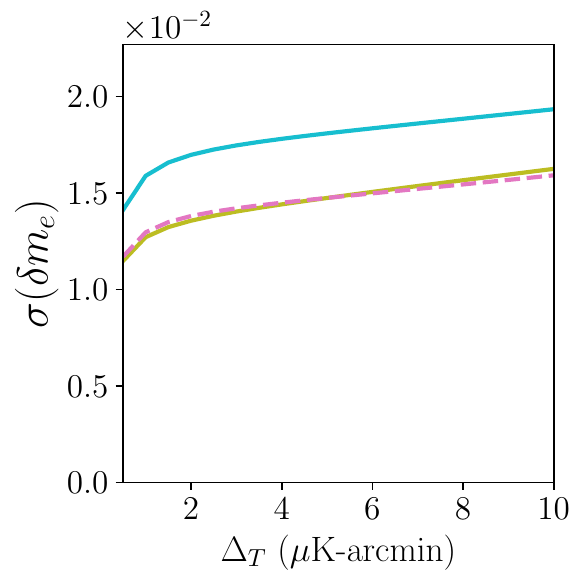}
    \includegraphics[height=4cm]{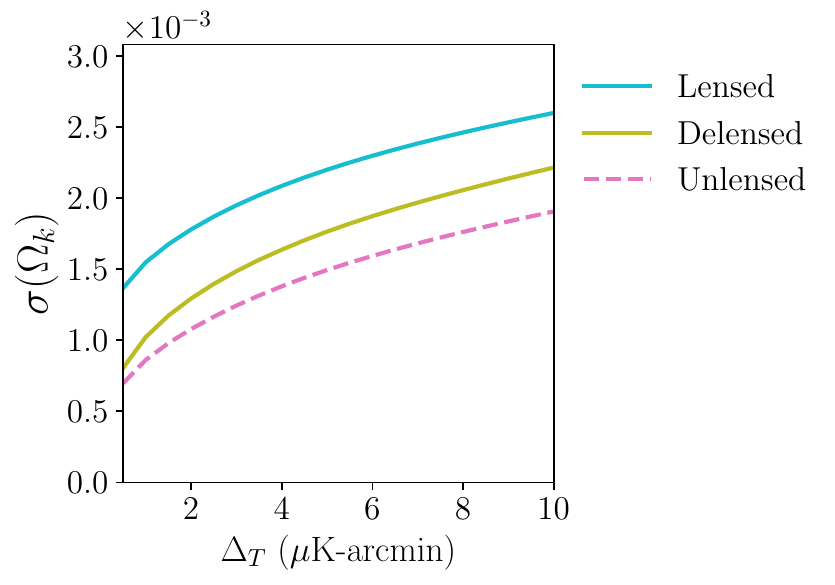}
    \caption{Forecasted constraints on $H_0$, $\delta m_e$, and $\Omega_k$ in an 8-parameter $\Lambda\text{CDM} + \delta m_e + \Omega_k$ cosmology. Note the lower delensed error from $\delta m_e$ and degenerate parameter $H_0$, as in Fig.~\ref{fig:constraints_varyingConstants}}
    \label{fig:constraints_varyingConstantsCurvature}
\end{figure}

In Ref.~\cite{Sekiguchi:2020teg}, a non-zero spatial curvature was also found to ease the Hubble Tension by shifting $D_A^*$ for the CMB. We utilize a fiducial value of $\Omega_k = -8.9719 \times 10^{-3}$, as guided by Ref.~\cite{Schoneberg:2021qvd}. In Fig.~\ref{fig:constraints_varyingConstantsCurvature} we show the results for 8-parameter forecasts with varying $\delta m_e$ and $\Omega_k$ in addition to $\Lambda$CDM parameters. From left to right, the panels show constraints for $H_0$, $\delta m_e$, and $\Omega_k$ respectively, as a function of noise. At decreasing noise levels, delensing provides a significant improvement in the constraints. The improvement from lensed to delensed constraints is shown in Table~\ref{tbl:improvement_varyingConstantsCurvature}. Note that the model which most significantly reduced the Hubble Tension in Ref.~\cite{Schoneberg:2021qvd}, $\Lambda\text{CDM} + \delta m_e + \delta \Omega_k$ has an improvement in the $H_0$ error from delensing of 18.0\%, which is roughly on par to reducing the noise of lensed observations from 10 to 1 $\unit{\mu K}$-arcmin, which would require a hundred-fold increase in the number of detector-years of observing. 

\begin{figure}[t!]
    \centering
    \includegraphics[width=0.4\columnwidth]{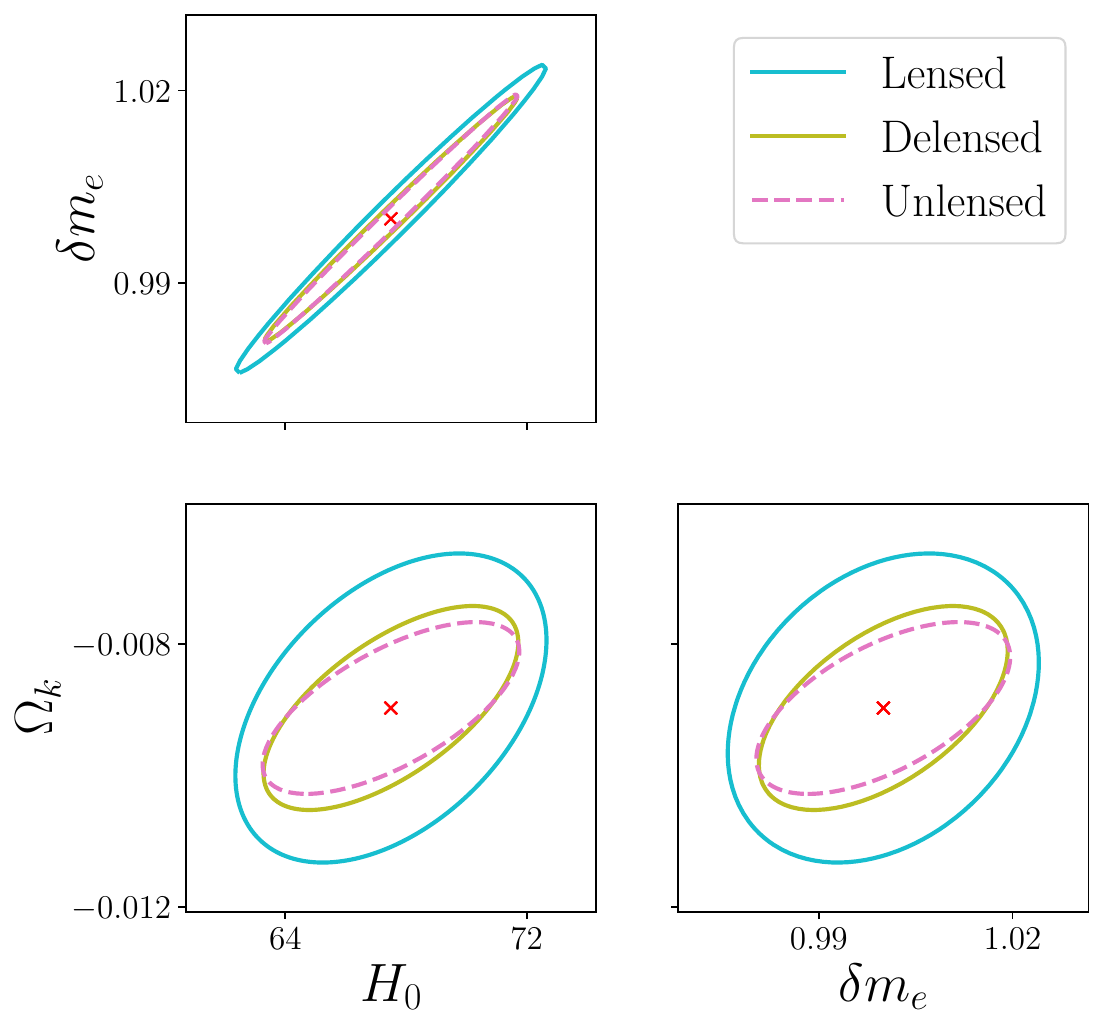}
    \caption{Forecasted errors and degeneracy of parameters of $\Lambda\text{CDM} + \delta m_e + \Omega_k$ cosmology at $1~\unit{\mu K}$-arcmin. 
    }
    \label{fig:contours_varyingConstantsCurvature}
\end{figure}

In Fig.~\ref{fig:contours_varyingConstantsCurvature}, we show the forecasted uncertainty and degeneracy of parameters of this $\Lambda\text{CDM} + \delta m_e + \Omega_k$ model (without BAO) before and after delensing. As with zero curvature models, delensing provides a significant tightening of constraint contours for all parameters of interest.

\begin{table}[t!]
\centering
\begin{booktabs}{
  colspec = {lcccccccc},
  cell{1}{2,6} = {c=4}{c}, 
  cell{1}{1} = {c=1}{c},
}
\toprule
  Model                                            &  With BAO & & &                                                   & Without BAO & & & \\
\midrule
                                                   &  $H_0$         & $\delta\alpha$ & $\delta m_e $  & $\Omega_k$     &  $H_0$         & $\delta\alpha$ & $\delta m_e $  & $\Omega_k$     \\
\cmidrule[lr]{2-5}\cmidrule[lr]{6-9}
$\Lambda\text{CDM}+\delta\alpha+\delta m_e+\Omega_k$& 12.4          &  10.5          &  14.4          &  14.7          &  18.7          &  11.5          &  20.3          &  19.2          \\
$\Lambda\text{CDM}+\delta\alpha+\delta m_e$         &  1.9          &  10.8          &  5.6           &                &  21.4          &  11.4          &  21.7          &               \\
$\Lambda\text{CDM}+\delta\alpha+\Omega_k$          &  4.8           &  10.4          &                &  5.9           &  19.9          &  11.0          &                &   20.6        \\
$\Lambda\text{CDM}+\delta\alpha$                   &  6.4           &  9.1           &                &                &  9.6           &  10.9          &                &               \\
$\Lambda\text{CDM}+\delta m_e+\Omega_k$            &  13.0          &                &   14.4         &   15.0         &  18.0          &                &   19.8         &  19.1         \\
$\Lambda\text{CDM}+\delta m_e$                     &  2.2           &                &   3.8          &                &  21.0          &                &   21.2         &               \\
$\Lambda\text{CDM}+\Omega_k$                       &  1.1           &                &                &   4.4          &  19.3          &                &                &  20.5         \\
$\Lambda\text{CDM}$                                &  2.6           &                &                &                &  4.9           &                &                &               \\
\bottomrule
\end{booktabs}
\caption{Table displaying percent improvement in error from delensing in the case of varying fundamental constants with non-zero curvature. This is computed with $100\times(1- \frac{\sigma_\text{del}}{\sigma_\text{len}})$ at $1~\unit{\mu K}$-arcmin (the approximate noise level of CMB-S4). Empty cells indicate parameters fixed at their fiducial values.}
\label{tbl:improvement_varyingConstantsCurvature}
\end{table}

\subsection{Early Dark Energy}
\label{subsec:ede}

\begin{figure}[t!]
    \centering
    \includegraphics[width=\columnwidth]{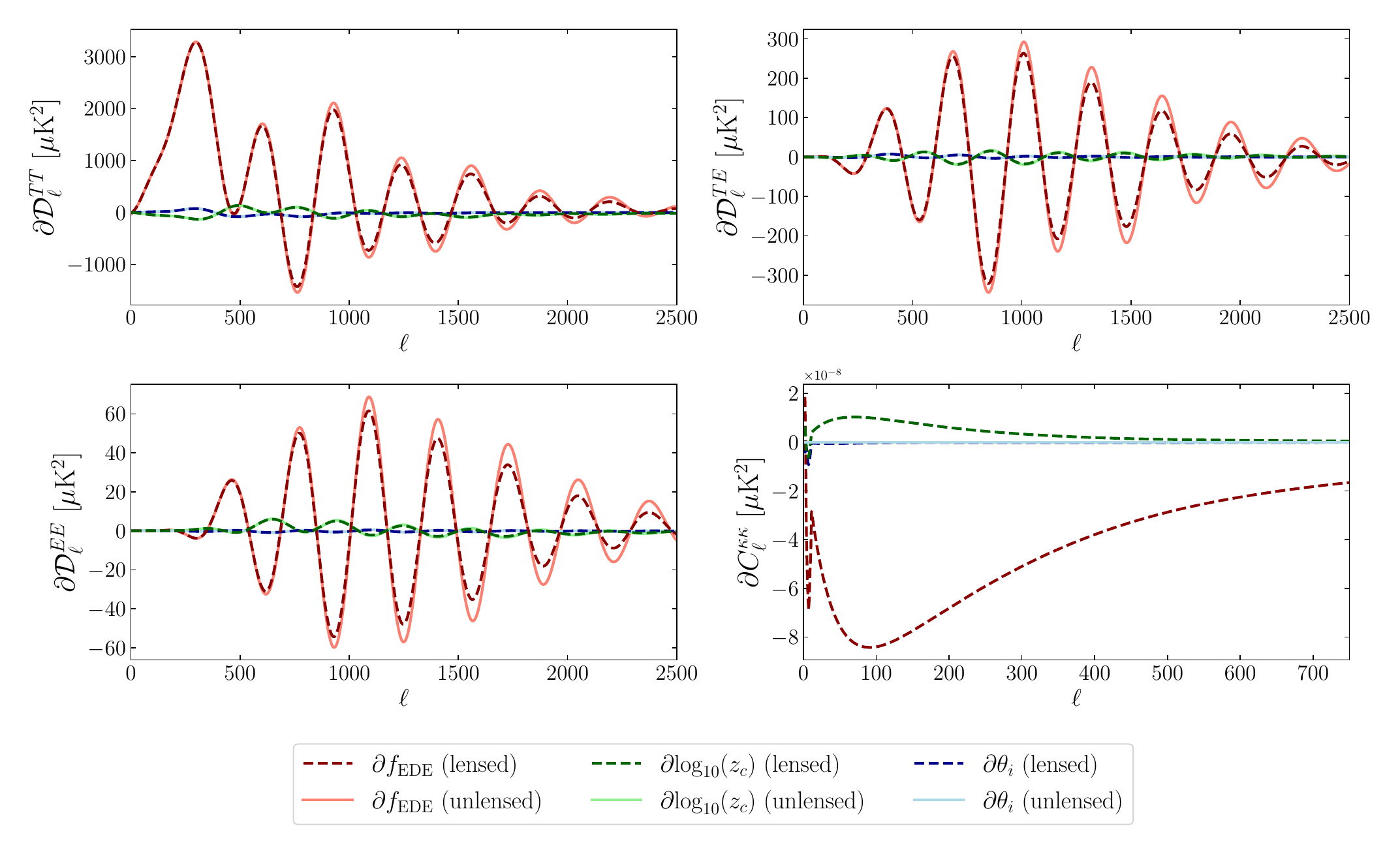}
    \caption{Derivatives of CMB power spectra ($TT$, $TE$, and $EE$) and lensing power spectrum ($\kappa\kappa$) with respect to effective early dark energy parameters: the fractional early dark energy density at the transition redshift $f_{\text{EDE}}$, the critical reshift when the early dark energy field becomes dynamical $\text{log}_{10}(z_c)$, and the initial axion misalignment angle $\theta_i$. Note the larger derivatives for unlensed compared to lensed spectra (especially for $f_{\text{EDE}}$), indicating that delensing could serve to better isolate CMB peaks and features associated with the parameters, just as in the case of varying fundamental constants.}
    \label{fig:spectraDeriv_EDE}
\end{figure}

Early dark energy refers to a class of models aimed at resolving the Hubble Tension by including a scalar field initially frozen-in due to Hubble friction that performs damped oscillations around its local minimum once the Hubble parameter drops below a critical value. Physically, this describes an axion-like field that once behaved like a cosmological constant, but began to decay at some critical redshift $z_c$ \cite{Karwal:2016vyq}. By increasing the Hubble parameter for a limited time in the early universe, the sound horizon and diffusion damping scale decrease, and field perturbations have additional pressure (which produces signature effects on CMB spectra, as seen in Fig.~\ref{fig:spectraDeriv_EDE}) \cite{Smith:2019ihp}. By involving an increase in the expansion rate around matter/radiation equality due to this $\Lambda$CDM extension, the Hubble Tension can be somewhat alleviated without spoiling the fit to Planck measurements \cite{Schoneberg:2021qvd}. 

For our analysis, we utilize the extremely light field and potential of the form $V(\phi) = m^2 f^2 (1 - \cos(\phi/f))^n + V_\Lambda$. Instead of particle physics parameters $f$ and $m$, we utilize the corresponding effective early dark energy parameters $\text{log}_{10}(z_c)$, the critical redshift at which the field becomes dynamical, and $f_{\text{EDE}}$, the fractional energy contribution of the field at said redshift. We additionally include parameter $\theta_i$, the initial axion misalignment angle. We utilize the fiducial best-fit values of $f_{\text{EDE}} = 0.098$, $\text{log}_{10}(z_c) = 3.63$, and $\theta_i = 2.58$ taken from Ref.~\cite{Hill:2020osr} that are found to best reduce the tension. 
We incorporated the treatment of early dark energy from Ref.~\cite{Hill:2020osr}\footnote{\url{https://github.com/mwt5345/class_ede}} into \texttt{CLASS\_delens}
to perform the forecasts presented here.

\begin{figure}[tp!]
    \centering
    \includegraphics[height=3.2cm]{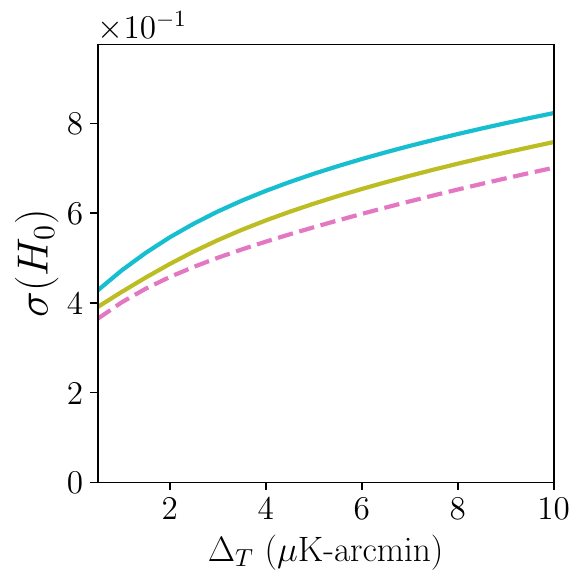}
    \includegraphics[height=3.2cm]{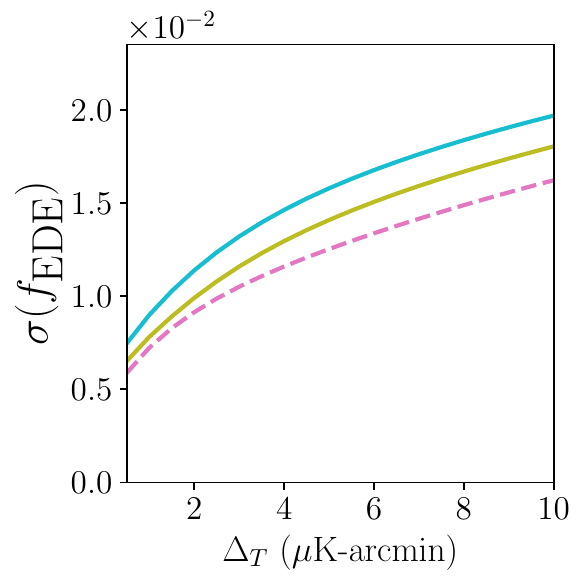}
    \includegraphics[height=3.2cm]{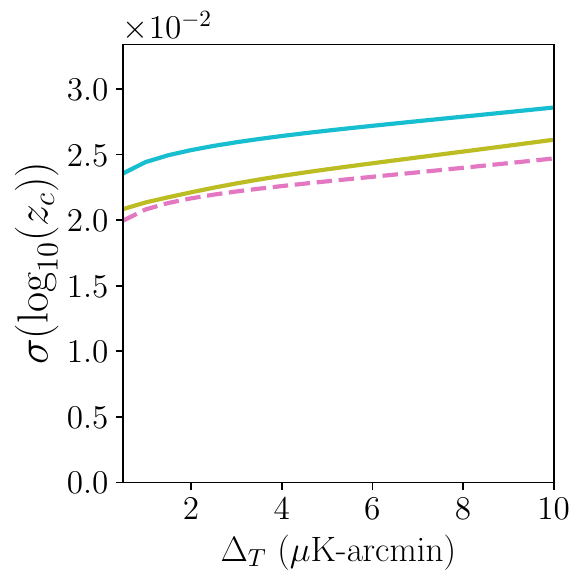}
    \includegraphics[height=3.2cm]{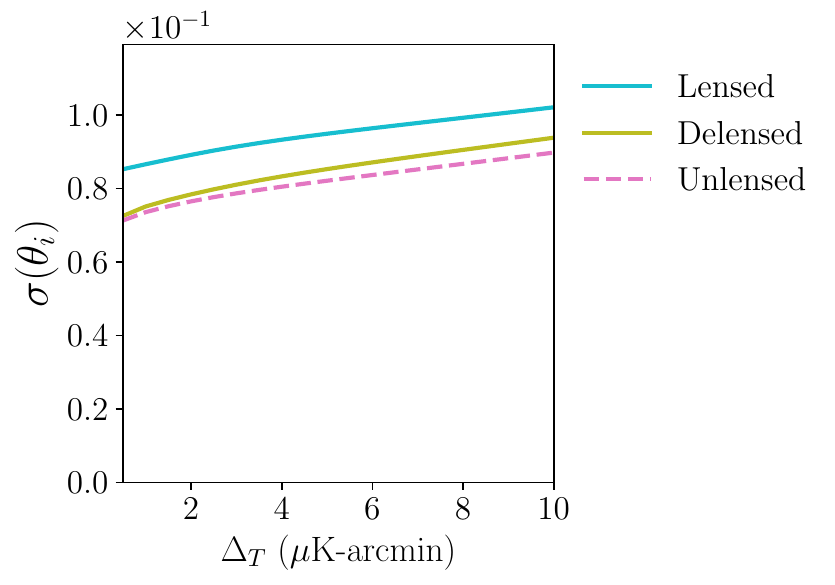}
    \caption{Forecasted constraints on $H_0$,  $f_{\text{EDE}}$,  $\text{log}_{10}(z_c)$, and  $\theta_i$ in a 9-parameter $\Lambda\text{CDM} + f_{\text{EDE}} + \text{log}_{10}(z_c) + \theta_i$ cosmology.
    }
    \label{fig:constraints_EDE}
\end{figure}

\begin{figure}[tp!]
    \centering
    \includegraphics[width=0.6\columnwidth]{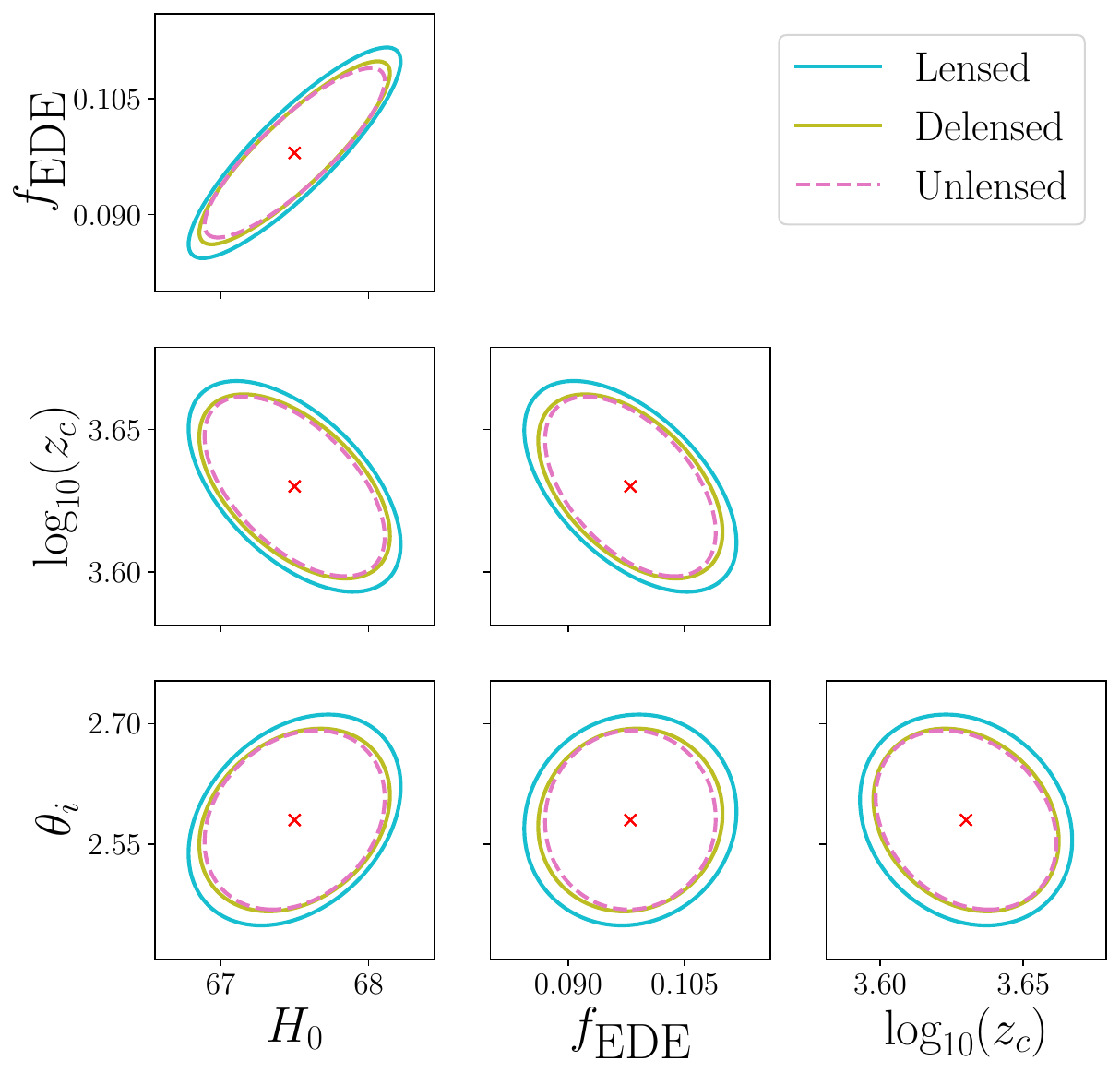}
    \caption{Forecasted errors and degeneracy of parameters of $\Lambda\text{CDM} + f_{\text{EDE}} + \text{log}_{10}(z_c) + \theta_i$ cosmology at $1~\unit{\mu K}$-arcmin. 
    }
    \label{fig:contours_EDE}
\end{figure}

In Fig.~\ref{fig:constraints_EDE} we show the results for $\Lambda\text{CDM} + f_{\text{EDE}} + \text{log}_{10}(z_c) + \theta_i$ cosmology forecasts. The percent improvement from lensed to delensed constraints is shown in Table~\ref{tbl:improvement_EDE}. In Fig.~\ref{fig:contours_EDE}, we show the forecasted uncertainty and degeneracy of parameters of the $\Lambda\text{CDM} + f_{\text{EDE}} + \text{log}_{10}(z_c) + \theta_i$ model (without BAO data) before and after delensing. Delensing provides a significant tightening of constraint contours for all parameters of interest.

\begin{table}[t!]
\centering
\begin{booktabs}{
  colspec = {lcccccccc},
  cell{1}{2,6} = {c=4}{c}, 
  cell{1}{1} = {c=1}{c},
}
\toprule
  Model                                              &  With BAO & & &                                                  & Without BAO & & & \\
\midrule
                                                     &  $H_0$        & $f_\text{EDE}$ & $\text{log}_{10}(z_c)$ & $\theta_i$& $H_0$ & $f_\text{EDE}$ & $\text{log}_{10}(z_c)$ & $\theta_i$  \\
\cmidrule[lr]{2-5}\cmidrule[lr]{6-9}
$\Lambda\text{CDM}+f_\text{EDE}+\text{log}_{10}(z_c)+\theta_i$& 8.1  & 9.3            & 10.7           & 7.8            & 9.5         & 10.5           & 10.7           & 9.9           \\
$\Lambda\text{CDM}+f_\text{EDE}+\text{log}_{10}(z_c)$&   8.3         & 9.4            & 10.9           &                & 8.1         & 9.8            & 10.8           &               \\
$\Lambda\text{CDM}+\text{log}_{10}(z_c)+\theta_i$    &  2.0          &                & 10.4           & 7.8            & 5.8         &                & 10.2           & 9.2           \\
$\Lambda\text{CDM}+\text{log}_{10}(z_c)$             &  2.0          &                & 10.5           &                & 4.1         &                & 10.4           &               \\
$\Lambda\text{CDM}+f_\text{EDE}+\theta_i$            &  7.9          &  9.0           &                & 8.0            & 9.3         & 10.0           &                & 10.0          \\
$\Lambda\text{CDM}+f_\text{EDE}$                     &  7.9          &  9.0           &                &                & 8.0         & 9.4            &                &               \\
$\Lambda\text{CDM}+\theta_i$                         &  2.2          &                &                & 8.0            & 6.1         &                &                & 9.4           \\
$\Lambda\text{CDM}$                                  &  2.0          &                &                &                & 4.3         &                &                &               \\
\bottomrule
\end{booktabs}
\caption{Table displaying percent improvement in error from delensing in the case of early dark energy. This is computed with $100\times(1- \frac{\sigma_\text{del}}{\sigma_\text{len}})$ at $1~\unit{\mu K}$-arcmin (the approximate noise level of CMB-S4). Empty cells indicate parameters fixed at their fiducial values.}
\label{tbl:improvement_EDE}
\end{table}

\subsection{Self-Interacting Dark Radiation}
\label{subsec:sidr}

\begin{figure}[t!]
    \centering
    \includegraphics[width=\columnwidth]{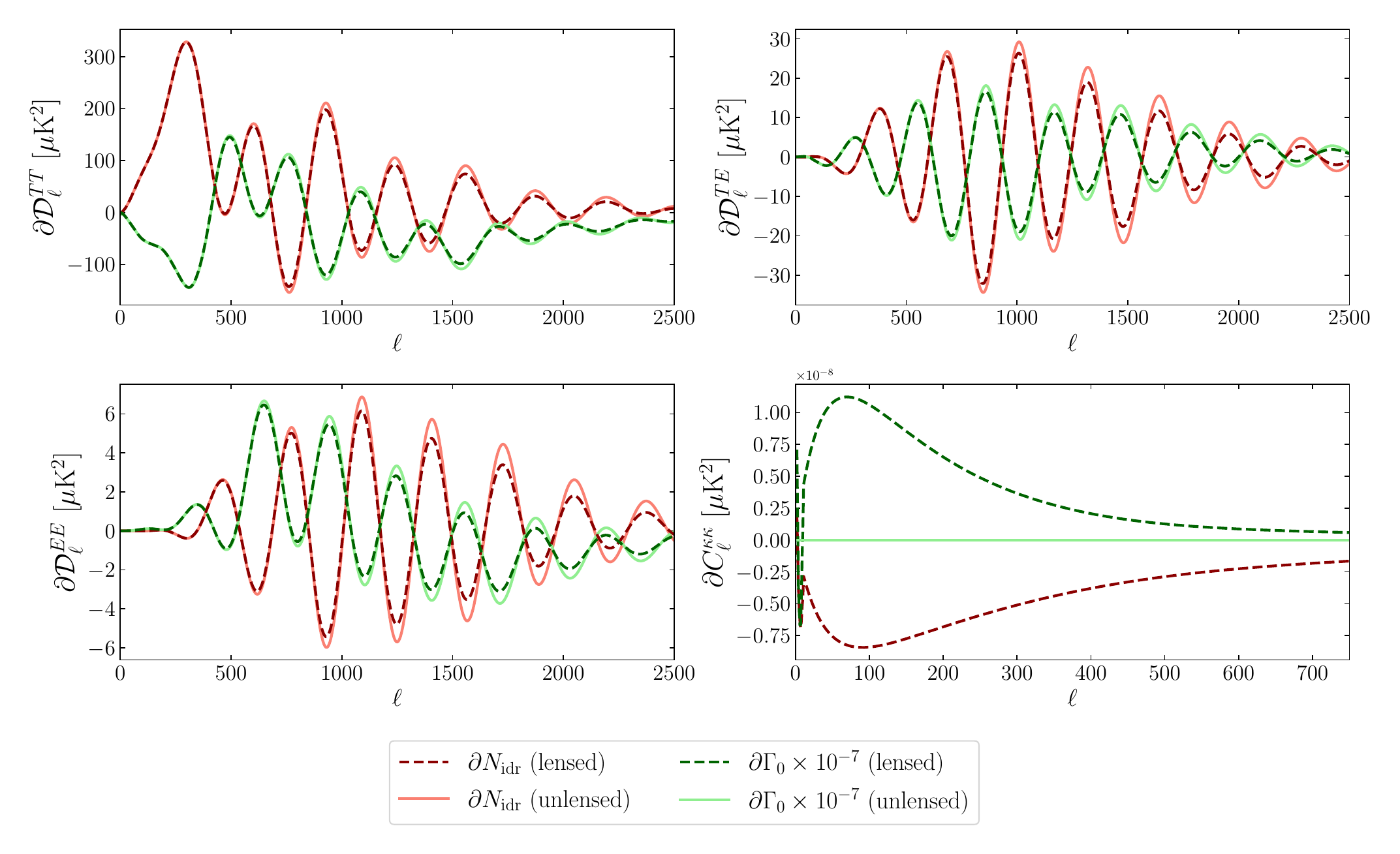}
    \caption{Derivatives of CMB power spectra ($TT$, $TE$, and $EE$) and lensing power spectrum ($\kappa\kappa$) with respect to the energy density in interacting dark radiation $N_\text{idr}$ and the present rate of momentum transfer between the dark matter and dark radiation $\Gamma_0$. Note the larger derivatives for unlensed compared to lensed spectra, indicating that delensing could serve to better isolate CMB peaks and features associated with the parameters.
    Here and below, $\Gamma_0$ is reported in $\unit{Mpc^{-1}}$.
    }
    \label{fig:spectraDeriv_SIDR}
\end{figure}

Some of the most well-known extensions to the $\Lambda$CDM model take the form of free-streaming massless relics (dark radiation). Such extensions are relatively well-constrained by the CMB \cite{Bashinsky:2003tk, Schoneberg:2021qvd}, but there is no physical necessity that these relics are free-streaming. While free-streaming and self-interacting species would contribute equally to the energy budget of the universe and CMB damping tail, only free-streaming particles induce the observed, characteristic phase shift in the anisotropies~\cite{Bashinsky:2003tk,Baumann:2015rya}, meaning there may be non-free-streaming particles at play that are not as well constrained. Such self-interacting species would form a relativistic fluid that changes the expansion rate during radiation domination, similarly to the early dark energy of Sec.~\ref{subsec:ede}, therefore influencing the size of the sound horizon at last scattering and shifting the inference of $H_0$ (see Fig.~\ref{fig:spectraDeriv_SIDR}).

We utilize the fiducial best-fit value of $N_{\text{idr}} = 0.3867$ from Ref.~\cite{Schoneberg:2021qvd}, where $N_{\text{idr}}$ is the contribution of self-interacting dark radiation to the effective neutrino number $N_{\text{eff}}$. The percent improvement from lensed to delensed constraints is shown in Table~\ref{tbl:improvement_SIDR}.


\begin{table}[t!]
\centering
\begin{booktabs}{
  colspec = {lcccc},
  cell{1}{2,4} = {c=2}{c}, 
  cell{1}{1} = {c=1}{c},
}
\toprule
  Model                                            &  With BAO &                      & Without BAO & \\
\midrule
                                                   &  $H_0$         & $N_\text{idr}$  &  $H_0$         & $N_\text{idr}$ \\
\cmidrule[lr]{2-3}\cmidrule[lr]{4-5}
$\Lambda\text{CDM} + N_\text{idr}$                 &  7.9           &  11.1           &  7.1           &  11.2          \\
$\Lambda\text{CDM}$                                &  2.0           &                 &  4.4           &                \\
\bottomrule
\end{booktabs}
\caption{Table displaying percent improvement in error from delensing in the case of self-interacting dark radiation. This is computed with $100\times(1- \frac{\sigma_\text{del}}{\sigma_\text{len}})$ at $1~\unit{\mu K}$-arcmin (the approximate noise level of CMB-S4). Empty cells indicate parameters fixed at their fiducial values.}
\label{tbl:improvement_SIDR}
\end{table}

\begin{figure}[t!]
    \centering
    \includegraphics[height=4cm]{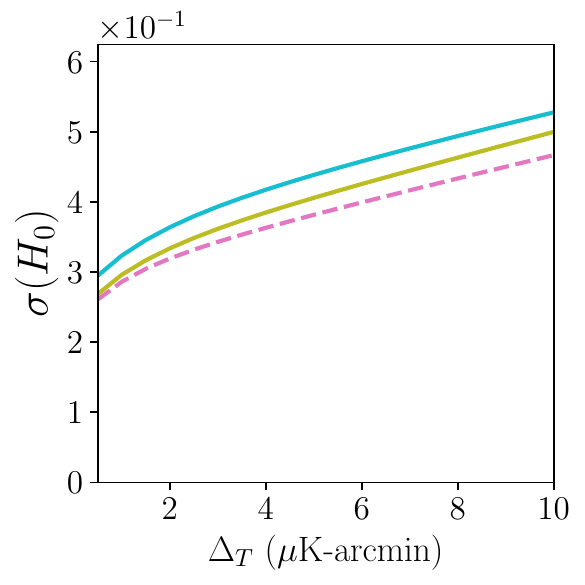}
    \includegraphics[height=4cm]{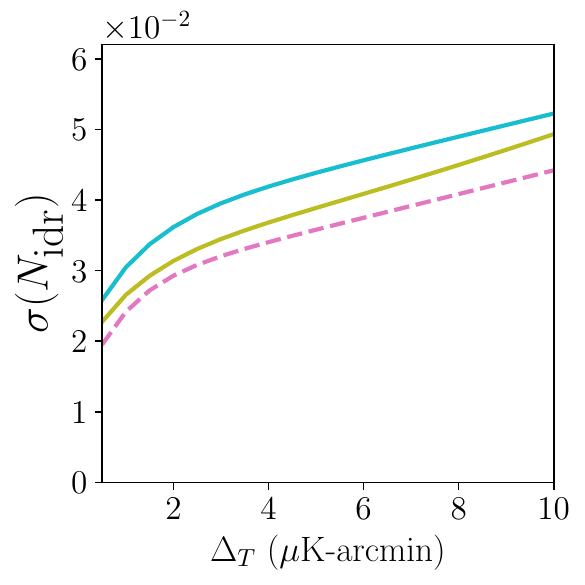}
    \includegraphics[height=4cm]{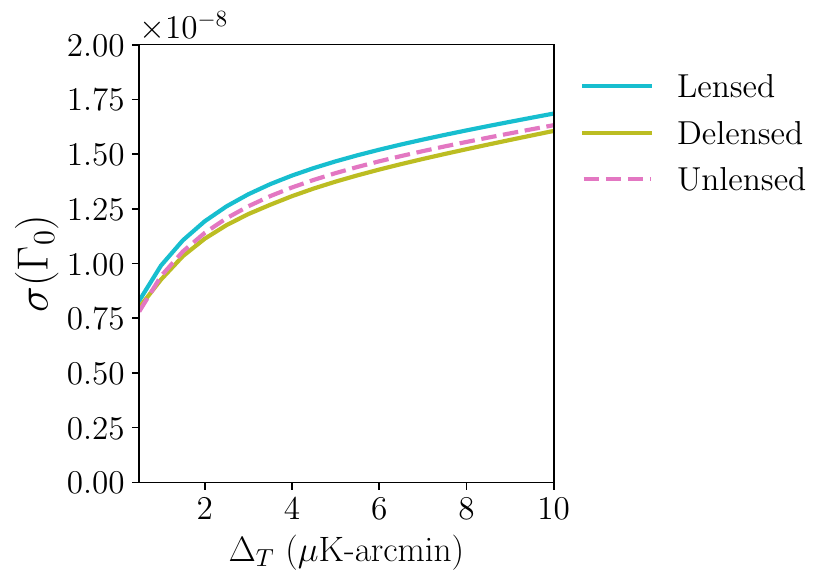}
    \caption{Forecasted constraints on $H_0$, $N_\text{idr}$, and $\Gamma_0$ in an 8-parameter $\Lambda\text{CDM} + N_\text{idr} + \Gamma_0$ cosmology.
    }
    \label{fig:constraints_SIDRDM}
\end{figure}

\subsubsection{Dark Radiation-Dark Matter Scattering}

\begin{figure}[t!]
    \centering
    \includegraphics[width=0.4\columnwidth]{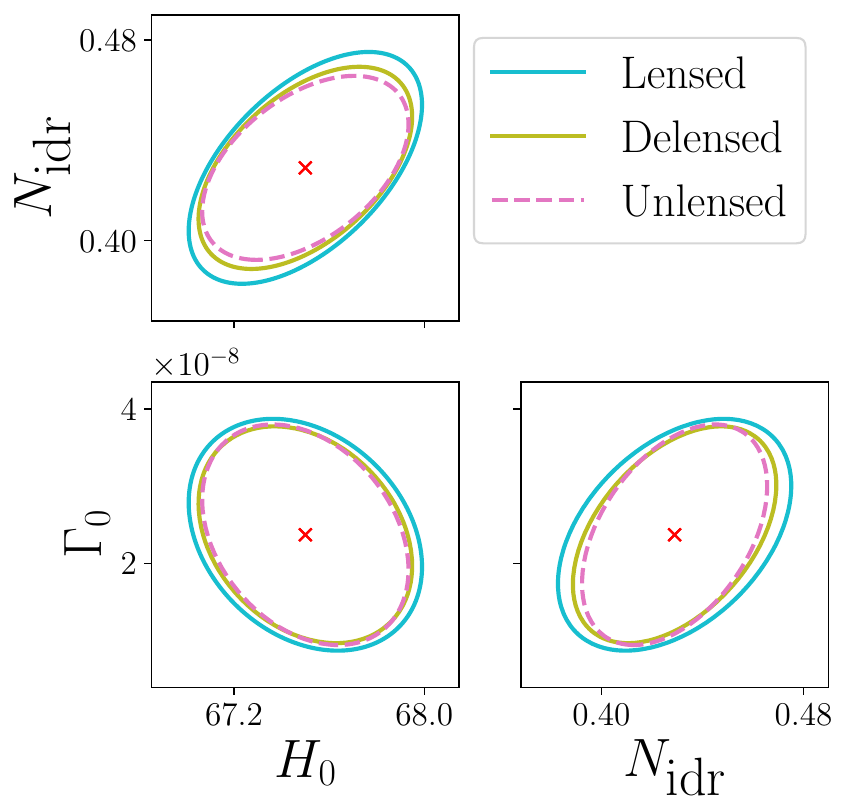}
    \caption{Forecasted errors and degeneracy of parameters of $\Lambda\text{CDM} + N_\text{idr} + \Gamma_0$ cosmology at $1~\unit{\mu K}$-arcmin. 
    }
    \label{fig:contours_SIDRDM}
\end{figure}

Additionally, it is possible that this self-interacting dark radiation futher interacts with dark matter. Such a cosmology is then additionally dependent on $\Gamma_0$, the present rate of momentum transfer between the dark matter and dark radiation \cite{Buen-Abad:2017gxg}. In this case, we utilize the fiducial best-fit values of $N_{\text{idr}} = 0.4290$ and $\Gamma_0=2.371\times10^{-8} ~\unit{Mpc^{-1}}$ from Ref.~\cite{Schoneberg:2021qvd}. Fig.~\ref{fig:constraints_SIDRDM} shows the results for the forecasts of $\Lambda$CDM and the 2-parameter extension of $N_{\text{idr}}$ and $\Gamma_0$. In Fig.~\ref{fig:contours_SIDRDM}, we show the forecasted uncertainty and degeneracy of parameters of the $\Lambda\text{CDM} + N_\text{idr} + \Gamma_0$ model (without BAO data) before and after delensing. The improvement from lensed to delensed constraints is shown in Table~\ref{tbl:improvement_SIDRDM}.

\begin{table}[t!]
\centering
\begin{booktabs}{
  colspec = {lcccccc},
  cell{1}{2,5} = {c=3}{c}, 
  cell{1}{1} = {c=1}{c},
}
\toprule
  Model                                            &  With BAO & &                      & Without BAO & & \\
\midrule
                                                   &  $H_0$         & $N_\text{idr}$  & $\Gamma_0$     &  $H_0$         & $N_\text{idr}$  & $\Gamma_0$     \\
\cmidrule[lr]{2-4}\cmidrule[lr]{5-7}
$\Lambda\text{CDM}+N_\text{idr}+\Gamma_0$          & 8.1            & 11.0            & 4.5            & 7.3            & 10.9            & 6.1            \\
$\Lambda\text{CDM}+N_\text{idr}$                   & 7.9            & 11.1            &                & 7.1            & 11.1            &                \\
$\Lambda\text{CDM}+\Gamma_0$                       & 1.9            &                 & 4.6            & 5.9            &                 & 6.3            \\
$\Lambda\text{CDM}$                                & 2.0            &                 &                & 4.4            &                 &                \\
\bottomrule
\end{booktabs}
\caption{Table displaying percent improvement in error from delensing in the case of self-interacting dark radiation with dark matter scattering. This is computed with $100\times(1- \frac{\sigma_\text{del}}{\sigma_\text{len}})$ at $1~\unit{\mu K}$-arcmin (the approximate noise level of CMB-S4). Empty cells indicate parameters fixed at their fiducial values.}
\label{tbl:improvement_SIDRDM}
\end{table}

\section{Conclusion}
\label{sec:conclusion}

The CMB is a key tool in understanding the foundational development of our universe. With higher precision measurements of CMB temperature and polarization anisotropies, more precise constraints for $\Lambda$CDM parameters have been made with remarkable levels of confidence \cite{Planck:2018vyg}. Discrepancies like the Hubble Tension may point the way toward novel physics in the early universe~\cite{Schoneberg:2021qvd}, and it is valuable to test these $\Lambda$CDM extensions with observations.  As more low-noise CMB data becomes available, delensing will be a valuable tool to increase the cosmological constraining power provided by our observations.

In this work, we provided a quantitative look at the improvements on constraints related to models aimed at addressing the Hubble Tension made possible by iterative delensing. We implemented three broad categories of early universe solutions (varying fundamental constants, early dark energy, and self-interacting dark radiation), estimated the effects of map-level delensing on generated power spectra, and forecasted constraints on model parameters. We demonstrated that constraints on $H_0$ derived from delensed spectra can be about 20\% tighter than from lensed spectra in the models we studied for upcoming CMB surveys like Simons Observatory and CMB-S4. Delensing provides significantly improved constraining power for model parameters across the board and is an analysis technique that does not require any instrumental design changes to implement with future data. CMB delensing proves an invaluable tool to glean as much information as possible about the early universe, regarding the Hubble Tension and beyond.


\acknowledgments

We thank Daniel Green, Selim Hotinli, Cynthia Trendafilova, and Alexander van Engelen for helpful conversations.
We would like to thank the Southern Methodist University (SMU) Physics Department for their continued support of research and foundation of education. We would like to thank the SMU Dedman College Interdisciplinary Institute and the Hamilton and Buford families for their funding and support over the course of this project through the Hamilton Scholars Program. 
This work was supported by the US~Department of Energy under Grant~\mbox{DE-SC0010129}.
Computational resources for this research were provided by SMU’s Center for Research Computing.





\bibliographystyle{utphys}
\bibliography{biblio.bib}

\end{document}